\documentclass[aps,prl,reprint,groupedaddress]{revtex4-2}

\usepackage{graphicx}
\usepackage{amsmath,amssymb,multirow}
\usepackage{bm}
\usepackage{braket}
\usepackage[
colorlinks,
linkcolor=blue,
urlcolor=blue,
citecolor=blue
]{hyperref}

\begin{document}

\title{Series of molecular-like doubly excited states of a quasi-three-body Coulomb system}

\author{M.\,G\'en\'evriez}
\email[]{matthieu.genevriez@uclouvain.be}
\author{M.\,Jungers}
\affiliation{Institute of Condensed Matter and Nanosciences, Universit\'e catholique de Louvain, BE-1348 Louvain-la-Neuve, Belgium}
\author{C.\,Rosen}
\author{U.\,Eichmann}
\email[]{eichmann@mbi-berlin.de}
\affiliation{Max-Born-Institute, 12489 Berlin, Germany}

\date{\today}

\begin{abstract}
We investigated resonant multiphoton excitation of high-angular-momentum doubly excited states of the strontium atom, a quasi-three-body Coulomb system, both experimentally and theoretically. A series of highly doubly excited states converging to the double ionization threshold was identified and reproduced by first-principles calculations, which provided access to the two-electron wavefunctions of the resonances. In these states, the outer electron is dynamically localized at large distances by pendulum-like oscillations of the inner electron across the nucleus. A complementary simple molecular-like adiabatic model offers a clear interpretation of the correlated two-electron motion and  suggests that the series is a general feature of highly doubly excited states of atoms and molecules.
\end{abstract}

\maketitle

Two highly excited electrons moving in the Coulomb field of a doubly charged
nucleus form a strongly correlated quantum system with a fascinatingly rich
spectrum of resonances~\cite{tanner00,aymar96}. As a prototype of the three-body
Coulomb problem, it stimulated intense experimental and theoretical work that
revealed the existence of longer-lived resonances in a two-electron phase
space that is otherwise largely chaotic~\cite{richter93}. In such states the
two electrons can follow remarkably correlated orbits~\cite{richter90,
eichmann90, genevriez21a}, such as the frozen-planet orbit, in which the two
electrons are dynamically localized on the same side of the nucleus at
specific orbit radii, the collinear eZe orbit, or the Langmuir
orbit~\cite{richter91,tanner00,muller93}.

Doubly-excited resonances have been experimentally detected as sharp lines in
photoexcitation spectra~\cite{domke91, czasch05, genevriez23, eichmann90} and,
with the advent of ultrafast lasers, their dynamics were unraveled in
the time domain~\cite{pisharody04, ott14, kaldun16}. To date, experimentally
identified series belong to those converging to a single ionization
threshold: one electron remains near the nucleus while the other explores
increasingly distant orbits, giving rise to a Rydberg-like progression of
asymmetric states~\cite{camus93,eichmann92,schulz96a,jiang05}. In contrast,
series converging to the double ionization threshold remain unexplored
experimentally despite abundant theoretical work~\cite{richter91,muller93,tanner00}. These
series mark a distinct, fascinating regime in which the independent-electron picture looses its meaning entirely and all three Coulomb interactions play an integral part in the correlated electron-pair dynamics~\cite{tanner00}.

In helium, series converging to the double ionization threshold have escaped
observation due to the small transition dipole moment from the ground state
and the excitation energies in the far ultraviolet
range~\cite{schulz96a,jiang05,jiang08}. Alkaline-earth atoms such as strontium
offer a more favorable platform because multiphoton resonant excitation makes
it possible to prepare doubly excited states inaccessible by one-photon
excitation from the ground state~\cite{camus89,eichmann90} with
high-resolution table-top
lasers~\cite{camus89,eichmann90,genevriez19,genevriez19b,teixeira20}. Their
two valence electrons, together with the doubly charged ion core, form
quasi-three-body Coulomb systems~\cite{aymar96} that have been crucial to
investigate planetary states~\cite{percival77,eichmann92}, i.e., asymmetric
doubly excited states for which the approximate individual principal quantum
numbers of the two electrons are different ($N < n$). Recent theoretical
effort~\cite{genevriez21,genevriez21a,genevriez23} has also made possible to
describe very high lying doubly-excited states of alkaline-earth atoms, well
beyond the degree of excitation reached in earlier pioneering
work~\cite{wood94}.

In this letter we report the experimental observation of a series of doubly
excited states converging to the double-ionization threshold. The series is
supported by a remarkable, correlated two-electron orbit which differs from
those theoretically identified in previous works: The inner electron
oscillates between opposite sides of the nucleus while the outer electron is
confined in a molecular-like potential at large distances. We thus propose to
refer to these states as pendular-planet states (PPS). The states are
identified in high-resolution spectra of doubly excited Sr that we recorded
and their assignment is supported by accurate calculations from first
principles. The PPS orbits are then identified using the wavefunctions
extracted from our calculations and the origin of the series is explained
using a simple adiabtic molecular-like model, which provides a clear physical
picture of this unsual series.

\begin{figure*}[t]
	\centering
	\includegraphics[width=\textwidth]{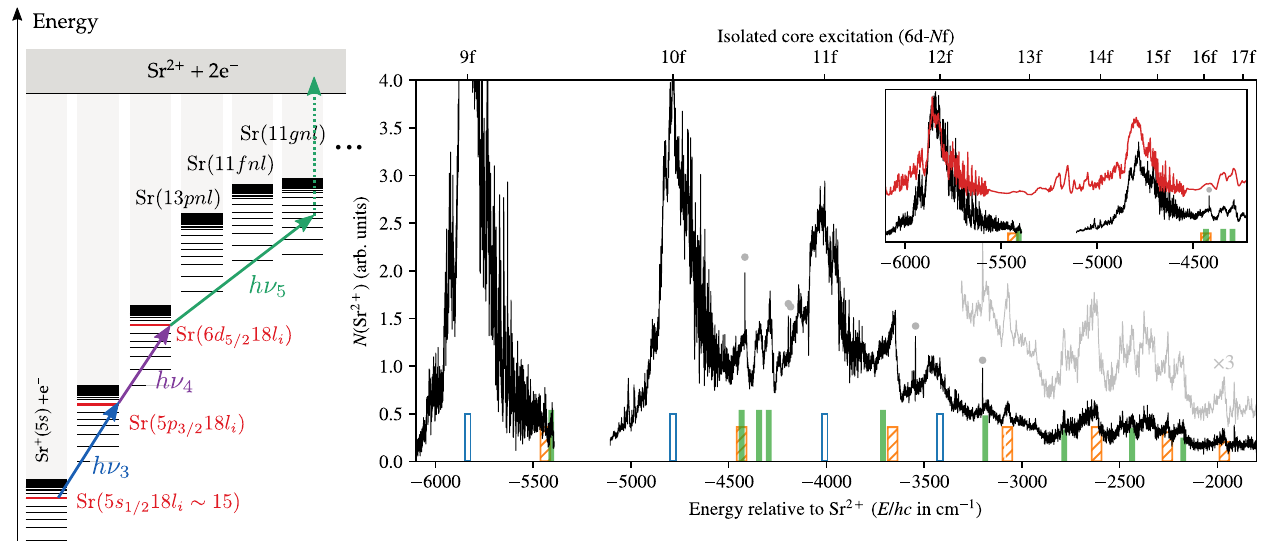}
	\caption{(Left) Resonant multiphoton excitation scheme. (Right) Experimental spectrum in the region of the Sr$^+(N=9-17)$ ionization thresholds. The wavenumber scale is given relative to
the Sr$^{2+}$ double ionization threshold to facilitate analysis. The gray
line shows the spectrum above -3800 cm$^{-1}$ scaled by 3 for visibility. The
energies of the Sr$^{+}(6dnl - Nfnl)$ transitions are shown in the upper
horizontal scale. The lines centered around those transitions and visible in
the spectrum are marked by vertical unfilled blue bars. The series of lines
under investigation is marked by the filled green bars. Hatched orange bars
denote a different series (see text). Spurious sharp resonances in the Sr$^+$
ion are marked by the full gray circles. The inset shows a comparison between
the experimental spectrum (lower full black line) and the theoretical spectrum (upper full red line) in the $N=9-10$ region.}
	\label{fig:overview_experimental_spectrum}
\end{figure*}

In the experiment, doubly excited states of Sr are prepared below the
Sr$^+(N=9-17)$ ionization thresholds following the well established sequential
resonant multiphoton excitation
scheme~\cite{eichmann89,eichmann92,rosen99,eichmann03} shown in
Fig.~\ref{fig:overview_experimental_spectrum}. Note that the
independent-particle quantum numbers adopted for the states classification are
only approximate because of the high degree of electronic correlation.
Briefly, Sr atoms in an effusive beam are first excited from the $5s^{2}$
ground state via an intermediate resonance to $5sn_il_i$ Rydberg state
($n_i=18$, $l_i = 15$) using two pulsed dye lasers and a ``Stark switching''
electric field~\cite{cooke78a,jones90} (see End Matter for details). After a
time delay of about 2~$\mu$s to adiabatically switch the Stark field to zero,
another two pulsed dye lasers excite the atoms to the $6d_{5/2}18l_i$ state
via the intermediate $5p_{3/2}18l_i$ state. Finally, a fifth pulsed dye laser
is scanned to excite the atoms further to doubly excited states near the
Sr$^+(N=9-17)$ thresholds. Approximately 200~ns after the fifth laser pulse, a
large electric field ($\sim 12$~kV/cm) is applied. It field ionizes atoms in
high lying doubly excited states and the highly excited Sr$^{+}$ ions produced
upon autoionization or upon photoionization by the fifth laser
pulse~\cite{rosen99,genevriez23}. Sr$^{2+}$ ions are recorded as a function of
the wavenumber of the fifth laser.

The experimental spectrum, shown in
Fig.~\ref{fig:overview_experimental_spectrum}, exhibits series of lines with a
rich substructure that are superimposed, at higher photon energies, onto a
continuum. The most prominent features, marked by unfilled blue bars, are centered
around the Sr$^+(6d-Nf)$ dipole-allowed transitions of the isolated ion.
Similar isolated-core-excitation (ICE) lines have been investigated at a lower
threshold ($N=7$)~\cite{heber97}, and the detailed analysis of the present
ones will be given elsewhere. Above $-3250$~cm$^{-1}$ the ICE series vanishes
but the spectrum is not structureless and two series of lines remain visible.
The first series, marked by the hatched orange bars, is associated to high lying
Rydberg-like states located below each hydrogenic ionization threshold
Sr$^+(N,L>3)$. Its analysis is ongoing. The second set of resonances, marked
by the filled green bars, is the PPS series investigated in the present paper. It is
remarkable because it remains visible even for very high excitation ($N=16$)
and because it converges to the double ionization threshold but does
\emph{not} follow the Rydberg formula or any other formulas describing the
energies of series of doubly-excited states (see~\cite{tanner00} for a
review).

To identify the nature of the PPS series we ran extensive theoretical
calculations using the method of configuration interaction with exterior
complex scaling (CI-ECS)~\cite{genevriez21}. It treats the two-electron motion
from first principles while accounting for the effect of the Sr$^{2+}$
closed-shell ion core with an accurate model potential. CI-ECS is capable of
describing highly doubly excited states of alkaline-earth
atoms~\cite{genevriez23,genevriez21a,genevriez19b,wehrli19}, a challenging
task as the correlated motion of the two electrons must be treated over large
radial distances ($> 1000\, a_0$), the density of states increases rapidly
($10^6$ eV$^{-1}$) and many decay channels are open. The calculations reported
here correspond to a degree of excitation ($N \sim 9-10$) that is well beyond
what had been previously calculated for any species other than helium
($N=6$)~\cite{wood94, genevriez21a}. Calculations details are given in the end
matter.

To assess the accuracy of the calculations, the photoionization spectrum
calculated with CI-ECS (see~\cite{genevriez21a} for details) is compared to
the experimental spectrum in the inset of
Fig.~\ref{fig:overview_experimental_spectrum}. To take into account the small
populations of states with $l_i \neq 15$ in the experiment (see end matter),
the theoretical spectra were calculated for several values of $l_i$ ($10-18$)
and subsequently added to reproduce as well as possible the experimental data.
The agreement between theory and experimental is excellent over the entire
wavenumber range and makes us confident that the theoretical calculations
capture all relevant details of the electron dynamics.

CI-ECS reveals that only a handful of resonances contribute to the PPS lines
in the experimental spectrum, typically one per value of the total
orbital-angular-momentum and spin quantum numbers. The analysis of the CI
coefficients further reveals that these states have predominant $Ndnl$
character, with $l=14$ for the resonance near $-5500$~cm$^{-1}$ and $l=16$ for
those near $-4500$~cm$^{-1}$. Their autoionization lifetimes fall in the range
from 100~fs to a few picoseconds.

\begin{figure}
	\centering
	\includegraphics[width=0.9\columnwidth]{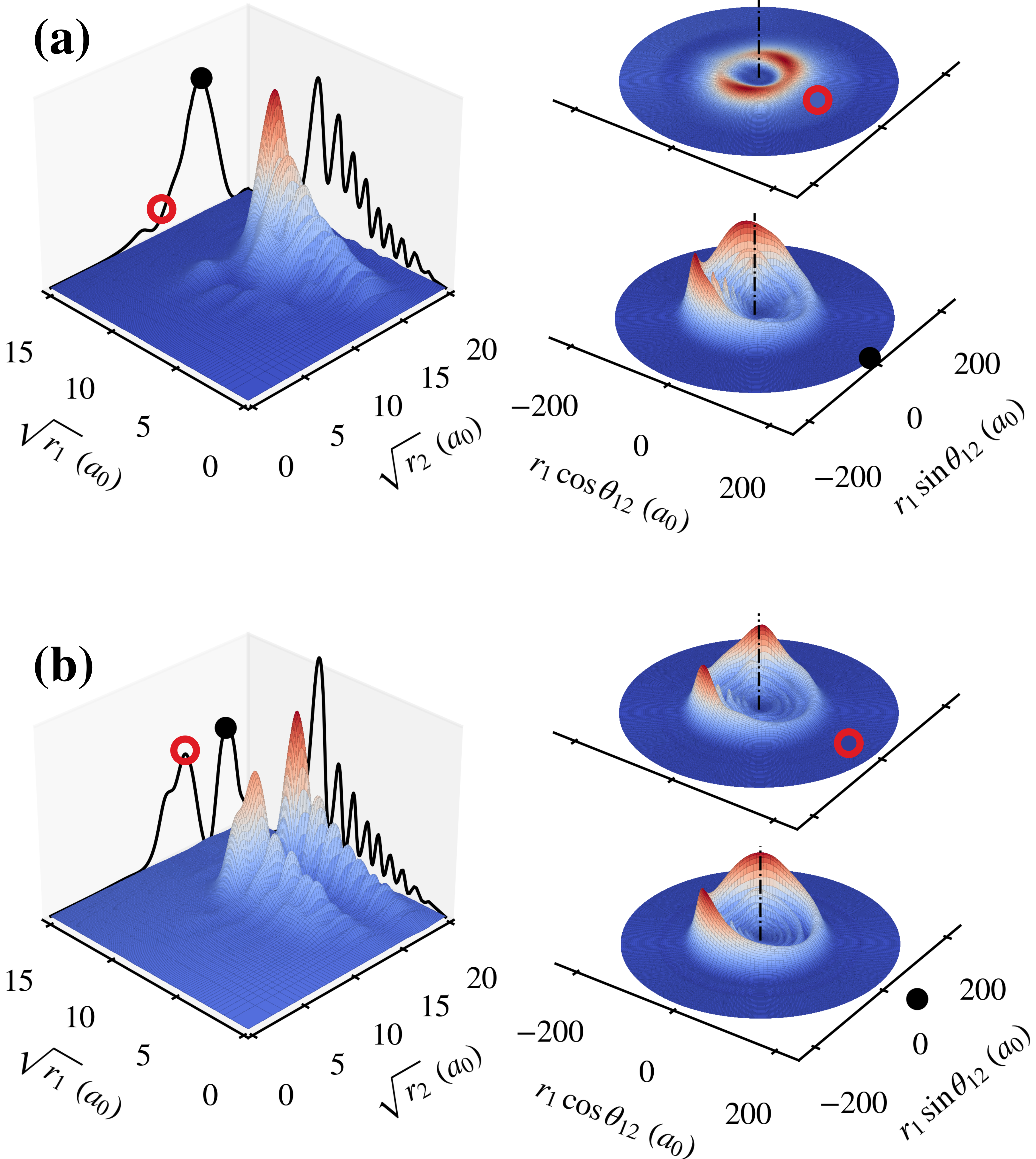}
	\caption{Two-dimensional representations of the two-electron probability density for states associated to the leftmost (a) and central (b) PPS resonances near $-4500$~cm$^{-1}$ (see text). The left panels correspond to the densities integrated over all spatial coordinates but $r_1$ and $r_2$. Projections onto the $r_1$ and $r_2$ axes are also shown. The right panels show the conditional density when the outer electron is fixed at a given radius, shown by the full black or red circles also shown on the left panels.}
	\label{fig:wavefunctions}
\end{figure}


Two-electron dynamics are revealed by inspecting the two-electron probability density $\left|\Psi(\bm{r}_1, \bm{r}_2)\right|^2$ along the radial coordinates $r_1$ and $r_2$ after integrating over all other spatial dimensions. They are shown in Fig.~\ref{fig:wavefunctions} (left column) for the the two leftmost resonances near $-4500$~cm$^{-1}$ with a total-angular-momentum quantum number of $L=17$. Note that the functions do not include antisymmetrization for clarity. Antisymmetrized densities are obtained by mirror symmetry about the $r_1 = r_2$ axis. 

The densities in the left column of Fig.~\ref{fig:wavefunctions} do not show a
strong dependency of the radial motion of one electron on the other. The
inner-electron density has $8$ minima along $r_1$, a value compatible with the
predominant $12d$ character of the inner-electron wavefunction deduced from
CI-ECS~\footnote{The quantum defect of the Sr$^+$ $Nd$ series is $\sim 1.5$}.
The outer-electron density of the leftmost state
[Fig.~\ref{fig:wavefunctions}(a)], on the other hand, shows no clear minima.
If electron correlations were neglected, this would be surprising as the
wavefunction of a highly excited electron has many nodes. The absence of nodes
thus suggests that the electronic repulsion plays an essential role and,
together with the Coulomb attraction of the Sr$^{2+}$ core and the centrifugal
barrier, gives rise to a local dynamical potential-energy well in which the
outer electron evolves. The state in Fig.~\ref{fig:wavefunctions}(a) may be
associated to the ground state of that potential, while the state associated
with the second resonance [Fig.~\ref{fig:wavefunctions}(b)], with one node, is
the first excited state. The departure of the wavefunctions of
Fig.~\ref{fig:wavefunctions} from what would be expected if electron
correlations were small, as in the case of planetary series observed in
earlier experiments, highlights the fundamentally different nature of the
PPS series.

Correlations in the angular motion of the two electrons are best observed by
fixing the outer electron at a given radius $r$ and plotting the conditional
density $\left|\Psi(r_1, \theta_{12} | r_2 = r)\right|^2$ (right panels in
Fig.~\ref{fig:wavefunctions}), integrated over all other spatial dimensions.
$\theta_{12}$ is the angle between the two electrons. When the outer electron
is near its most probable position, angular correlations are strong and the
two electrons reside predominantly on opposite sides of the nucleus. As the
outer electron gets closer to the nucleus [Fig.~\ref{fig:wavefunctions}(a),
top right graph], angular correlations drastically change, a fact to which we
shall come back later. Similar observations can be made about the probability
densities of other members of the PPS series and an example is shown in the
end matter.

\begin{figure}[ht]
	\centering
	\includegraphics[width=\columnwidth]{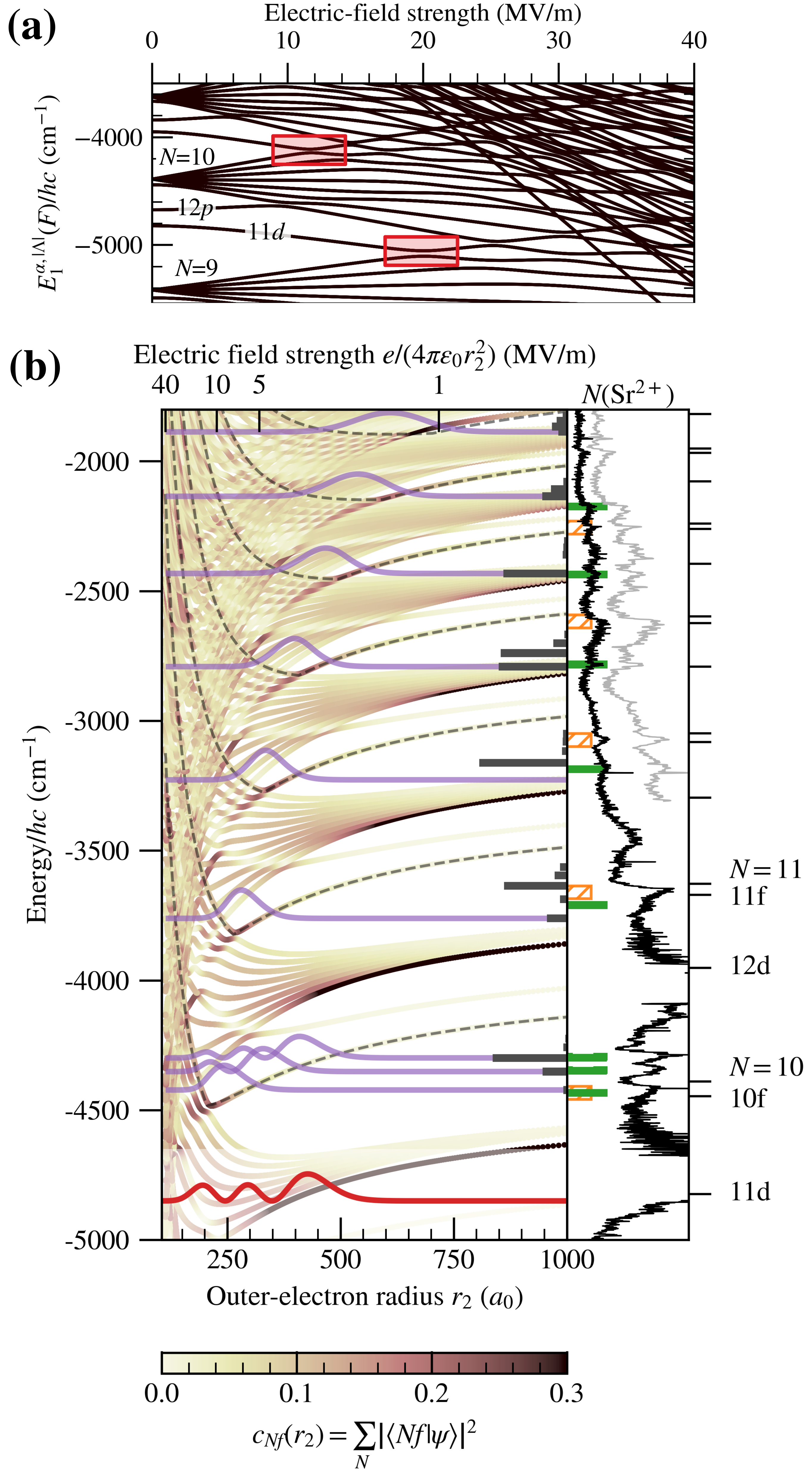}
	\caption{(a) Stark energies of Sr$^+$ in the $N=9-11$ region. (b) Adiabatic potentials $V(r_2)$. The colorscale shows the $Nf$ character of the inner-electron wavefunction. The radial probability densities $\left|\psi(r_2)\right|^2$ of the outer electron in the $6d18l_i$ state and in low-lying states of the potentials shown by the dashed lines are displayed as the red and purple lines, respectively. The horizontal black bars show the energy levels of the dashed-line potentials and their lengths give the transition strengths from the initial state. The right panel shows the experimental spectrum as in Fig.~\ref{fig:overview_experimental_spectrum}.}
	\label{fig:adiabatic_stark_potential}
\end{figure}

CI-ECS calculations successfully reproduce the PPS resonances observed in the
experiment and predict the two-electron dynamics, but do not give an intuitive
picture as to their physical origin. To do so we turn to a simple model in
which the fast motion of the inner electron is adiabatically decoupled from
the one of the (comparatively slower) outer electron. This is akin to a
Born-Oppenheimer approximation~\cite{heber97,feagin88} in which $r_2$ would be
the internuclear distance (see End Matter for details). The outer electron
generates an electric field which we assume to be uniform in the region where
the inner electron evolves and of strength $-1/r_2^2$ in atomic units. Because
of that field the inner electron undergoes a Stark shift that depends on
$r_2$, and we write its energy as $E_1^{\alpha, |\Lambda|}(-1/r_2^2)$. $\Lambda$ is the projection of
$\bm{L}$ onto the adiabtic axis $\bm{r}_2$, and $\alpha$ denotes all other
quantum numbers required to label the inner-electron Stark state. The inner-electron Stark energies are depicted in Fig.~\ref{fig:adiabatic_stark_potential}(a) as a function of the electric-field strength. Within these
approximations, the outer electron evolves in an adiabatic potential $V(r_2)$
that is the sum of the screened Coulomb attraction of the core ($-1/r_2$), the
centrifugal barrier ($l_2(l_2+1)/2r_2^2$), and the dynamically-Stark-shifted
energy $E_1^{\alpha,|\Lambda|}$ of the inner electron,
\begin{equation}
	V(r_2) = -\frac{1}{r_2} + \frac{l_2(l_2+1)}{2r_2^2} + E_1^{\alpha,|\Lambda|}(-1/r_2^2) .
	\label{eq:stark_potential}
\end{equation}
For the calculations shown below we take $l_2 = 16$ and $|\Lambda|=1$ based on
the analysis of the CI coefficients and wavefunctions given above.

Whereas the screened Coulomb attraction and the centrifugal barrier are
monotonous with $r_2$, the Stark energies $E_1^{\alpha,|\Lambda|}$ are not due
to avoided crossings between states [see
Fig.~\ref{fig:adiabatic_stark_potential}(a)]. As a result the adiabatic
potentials $V(r_2)$, shown in Fig.~\ref{fig:adiabatic_stark_potential}(b), may
exhibit local extrema. A local minimum is clearly visible for those potentials
for which the inner electron is predominantly in the Sr$^+(Nd)$ state at
large distances (dashed gray lines). It arises from the avoided crossings
shown by the red shaded areas in Fig.~\ref{fig:adiabatic_stark_potential}(a),
which involve the Stark states associated, at zero field, to the Sr$^+(Nd)$
state and the uppermost state from the Sr$^+(N-2L)$ hydrogenic manifold.
Remarkably, the lowest levels of the potential wells (purple lines) match the
energies of the PPS resonances (filled green bars, right panel) within less than 1\%.
For these calculation the small avoided crossings in the repulsive part of the
potentials were assumed to be traversed diabatically.

The model also explains why the PPS resonances are well visible in the
spectra. At the minimum of the potential wells the inner-electron state has a
strong $Nf$ character $c_{Nf}(r_2)$ [see colorscale in
Fig.~\ref{fig:adiabatic_stark_potential}(b)]. The transition electric-dipole
moment from the $6dn_il_i$ initial state is thus large when the outer electron
is near the bottom of the well. It is the case for low-lying states but, as
the degree of excitation of the outer electron increases, the wavefunctions
spread out and the transition moment decreases. This is illustrated by the
horizontal black bars in Fig.~\ref{fig:adiabatic_stark_potential}(b), whose
lengths are proportional to the effective, scaled transition moment
$\int_0^\infty \mathrm{d}r_2 \psi_2^*(r_2) c_{Nf}(r_2) d_{6d-Nf}N^{3/2}
\psi_{n_il_i}(r_2)$, with $d_{6d-Nf}$ the transition dipole moment of the
isolated ion and $\psi_{n_il_i}(r_2)$ and $\psi_2(r_2)$ the radial
wavefunctions of the outer electron in the initial and final states,
respectively.

The correlated two-electron motion encoded in the densities of
Fig.~\ref{fig:wavefunctions} is also in agreement with, and clarified by, the
model. The most probable radius for the outer electron in the ground state of
the potential near $-4500$~cm$^{-1}$ is $r^\text{max}_2 = 243~a_0$, in
excellent agreement with the one extracted from CI-ECS [$237~a_0$,
Fig.~\ref{fig:wavefunctions}(a)]. In both calculations the outer-electron
density has no nodes while the first excited state shows one node at $r_2 \sim
250~a_0$ [Fig.~\ref{fig:wavefunctions}(b)]. For sufficiently large $r_2$ the
inner-electron wavefunction has a predominant $Nd$ character and 8 nodes, as
expected. In the repulsive part of the potential, the dominant $Nd$ character
is however lost and the inner-electron acquires some character associated to
the uppermost states of the $(N-2)$ Stark manifold
[Fig.~\ref{fig:adiabatic_stark_potential}(a)]. In such states the
electron localizes on the side of the nucleus given by the direction of the
electric field~\cite{gallagher94} and the same effect is observed in the
two-electron density of Fig.~\ref{fig:wavefunctions}(a). For a small $r_2$
value (red circle), the inner electron is indeed predominantly on the same side of
the nucleus as the outer electron. The two-electron motion can thus be
semi-classically interpreted as follows: as the the outer electron oscillates
in its potential well the dipole moment of the ``inner electron + ion core''
system undergoes a periodic pendulum-like motion such that, when $r_2$ is
small, it effectively prevents the outer electron from penetrating into the
inner region. From the lifetime of the state ($\tau\sim 300$~fs) and the
spacing between adjacent energy levels in the well ($\sim 100$~cm$^{-1}$) we
deduce that the inner electron undergoes about 6 such flip-flops before the
system spontaneously decays by autoionization.

\textit{Conclusions} - We have identified a series of molecular-like doubly
excited states of a quasi three-body Coulomb system converging to the double
ionization both theoretically and experimentally. The states, which we call
pendular-planet states because of the underlying two-electron motion, find
their physical origins in avoided crossings between Stark states, a phenomenon
that occurs in all systems except hydrogen-like atoms. For this reason, we
expect the series to be a general feature of the doubly excited states of
atoms and molecules. Remarkably, a similar mechanism is responsible for the
long-range molecular binding between an ion and a Rydberg
atom~\cite{deiss21,duspayev21,zuber22}. In the latter case the electric field
is generated by an ion in the vicinity of the Rydberg atom whereas, in the
present case, the electric field is the one of the outer electron that
surrounds the Sr$^+$ ion, itself in a Rydberg state. Both cases illustrate
that the quasi-three-body Coulomb problem possesses series of resonances that,
because its third body is not so well defined~\cite{codling90}, are different
from the exact counterpart. Observing and characterizing the states discussed
above in other atomic and molecular systems is an exciting perspective for
future work that could be facilitated, in particular in more complex systems,
by the use of the simple model and physical picture we developed.

\begin{acknowledgements}
	This work was supported by the Fonds de la Recherche Scientifique under MIS Grant No. F.4027.24 and IISN Grant No. 4.4504.10. Computational resources were provided by the supercomputing facilities of the Université catholique de Louvain (CISM/UCL) and the Consortium des Équipements de Calcul Intensif en Fédération Wallonie Bruxelles (CÉCI) funded by the Fond de la Recherche Scientifique de Belgique (F.R.S.-FNRS) under convention 2.5020.11 and by the Walloon Region.
\end{acknowledgements}

\bibliographystyle{apsrev4-2}
%

\pagebreak
\phantom{a}
\pagebreak

\appendix
\section{End matter}

\subsection{Experimental details}

\textit{Stark switching.} In the experiment, $5sn_il_i$ Rydberg states with
large orbital angular momenta ($l_i \sim 15$) are produced by the ``Stark
switching'' technique~\cite{cooke78a,jones90}. Two pulsed dye lasers pumped by
an excimer laser first excite Sr atoms in their $5s^{2}$ ground state via an
intermediate resonance to a $5sn_ik_i$ Rydberg Stark state in the presence of
a static electric field. The field is then slowly
turned off ($\sim$ 2$\mu$s) such that the Stark state adiabtically evolves into a $5sn_il_i$
Rydberg state, the value of $l_i$ being determined by the value of $k_i$. In
the present case we choose $l_i = 15$. Nonadiabatic effects as the field is
switched off and stray electric fields in the chamber can cause the transfer
of a small part of the population to other, neighbouring $l$
values~\cite{genevriez21a,camus93}.

\textit{Sr$^{2+}$ production. } During the 5-photon excitation to highly doubly excited states, the pulse energies of the first four lasers are kept low ($\sim 10$~$\mu$J) and the fifth laser pulse is more intense ($\sim 1$~mJ). The latter is intense enough to photoionize the high lying doubly excited states, leaving Sr$^+$ in high Rydberg states~\cite{rosen99}. Autoionization of the doubly excited states can also produce highly excited Sr$^+$ ions. The $\sim 12$-kV/cm electric field pulse, applied 200~ns after the fifth laser pulse, field ionizes the highly excited Sr$^+$ ions and doubly-exicted Sr atoms produced by the lasers, yielding the Sr$^{2+}$ ions detected in the experiment. 

\subsection{Configuration interaction with exterior complex scaling}\label{sec:ci-ecs}
The method of configuration interaction with exterior complex
scaling (CI-ECS)~\cite{genevriez21} aims at accurately describing highly
doubly excited states of alkaline-earth atoms~\cite{genevriez23,genevriez21a,genevriez19b,wehrli19}. It treats the motion of
the two highly excited electrons of Sr explicitely while the
influence of the electrons of the Sr$^{2+}$ closed-shell core is accounted for with an empirical model
potential $V_l(r)$~\cite{genevriez21,aymar96}, yielding the following two-electron Hamiltonian
\begin{equation}
 H = -\frac12\bm{\nabla}_1^2 - \frac12\bm{\nabla}_2^2 + V_{l_1}(r_1) + V_{l_2}(r_2) + V_{l_1l_2}(\bm{r}_1, \bm{r}_2) + \frac{1}{r_{12}} . \label{eq:two-electron_hamiltonian}
\end{equation}
$\bm{r}_1$ and $\bm{r}_2$ are the coordinates of the electrons with respect to the nucleus and $r_{12} = |\bm{r}_1 - \bm{r}_2|$. The potentials $V_l(r)$ include monoelectronic core-polarization terms while the term $V_{l_1l_2}(\bm{r}_1, \bm{r}_2)$ accounts for dielectronic core polarization~\cite{genevriez21}. The latter is included for completeness only as its influence on the energies of the states presented above is negligible compared to experimental resolution ($\sim$ 1~cm$^{-1}$). The potentials $V_l(r)$ yield energies of Sr$^+$ accurate to better than a few cm$^{-1}$~\cite{genevriez21a,genevriez19b}. 

The two-electron Schrödinger equation is
solved using a basis of antisymmetrized products of one-electron spin-orbitals whose radial parts are numerical functions obtained with a finite-element discrete-variable-representation technique~\cite{rescigno00}. We use exterior complex scaling~\cite{nicolaides78,
simon79} to treat resonances and continuum processes with square integrable
functions, i.e., to be able to carry out the calculations in a box of finite
size ($1420\, a_0$). In
the present calculations, 80\,000 two-electron basis functions are used. The outer electron is described with a large set of spin-orbitals,
including in particular accurate Rydberg states up to $n=25$. The basis set
describing the inner electron is truncated to all states up to the Sr$^+(N=12)$ hydrogenic manifold. Angular momenta are $LS$ coupled and the
spin-orbit interaction is neglected because both electrons are highly excited
and thus predominantly far from the nucleus. The large, complex-symmetric
Hamiltonian matrix built from the two-electron functions is diagonalized in
the energy range of interest with a complex-symmetric Lanczos algorithm, yielding the complex-scaled wavefunctions and energies of bound
states, resonances and continuum states. We have checked that the resonances
energies and widths below the $N=11$ threshold are converged to better than a few cm$^{-1}$ with respect to the basis set size, complex-scaling radius and angle.

Electron dynamics are visualized as in Fig.~\ref{fig:wavefunctions} starting from the
wavefunctions $\Psi(\bm{r}_1, \bm{r}_2)$ calculated with CI-ECS. To reduce the
number of dimensions we integrate over the three Euler angles that define the orientation
of the electron--electron--Sr$^{2+}$-ion plane in
space~\cite{gremaud98,genevriez21a} to obtain the reduced two-electron probability density
\begin{align}
	\rho(r_1&, r_2, \cos\theta_{12}) =\\
	&\braket{\Psi | \delta(r_1 - r'_1)\delta(r_2 - r'_2)\delta(\cos\theta_{12} - \cos \theta'_{12}) | \Psi} , \nonumber
\end{align}
where $\delta(x)$ are Dirac delta functions. $\delta(\cos\theta_{12} - \cos \theta'_{12})$ is expanded in terms of Legendre polynomials~\cite{warner80} to evaluate the matrix element. $\rho$ depends only on the two electronic radial distances $r_1$
and $r_2$ and the interelectronic angle $\theta_{12}$, and can be readily used to generate the densities shown in Fig.~\ref{fig:wavefunctions}.

\subsection{Calculations at a higher threshold}

The size of the full CI-ECS basis set increases very rapidly close to the
double-ionization threshold, making converged calculations for Sr$^+(N>10)$
computationally too heavy. The analysis carried out in the main text however suggests that, to compute the resonances under
consideration, it is sufficient to consider inner-electron states associated to
the Sr$^+(N-2)$ and Sr$^+(N-1)$ hydrogenic manifold, and all states in between
with nonzero quantum defects. Such an ensemble is indeed sufficient
to reproduce the avoided crossing responsible for the potential well and the
strong $Nf$ character at its bottom.

\begin{figure}[ht]
	\centering
	\includegraphics[width=0.8\columnwidth]{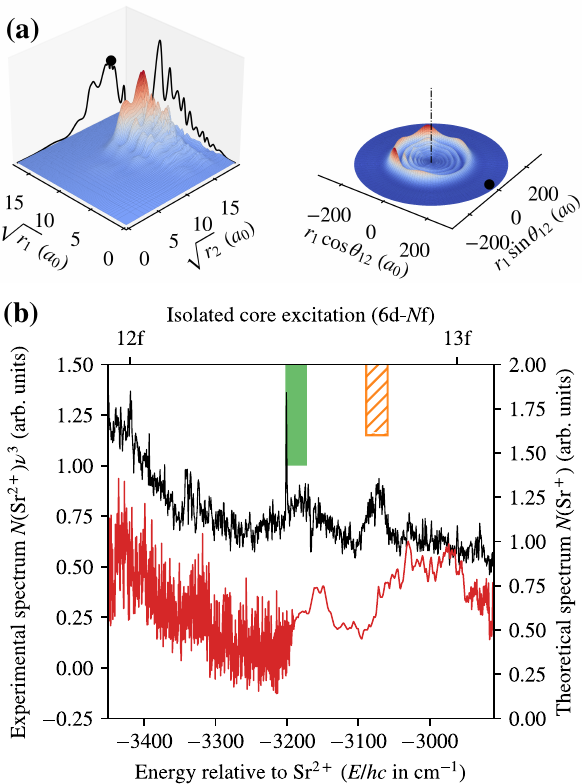}
	\caption{(a) Same as Fig.~\ref{fig:wavefunctions} for the resonance at $-3200$~cm$^{-1}$ [filled green bar in panel (b)] with a total-orbital-angular-momentum quantum number $L=13$. (b) Experimental (red) and theoretical (black) spectra in the region of the Sr$^+(N=12-13)$ ionization thresholds. Figure details are the same as those of Fig.~\ref{fig:overview_experimental_spectrum} (right panel).}
	\label{fig:N12-13_spectra}
\end{figure}

To verify this we ran a CI-ECS calculation in which the set of basis functions
describing the inner electron is limited to all states between, and including,
the $N=12$ and $N=13$ manifolds. The truncation of the basis set means that
most decay channels are neglected, and that the calculation is not fully
converged. We do not expect to reach quantitative agreement with experiment
but aim at providing a qualitative confirmation that the essential
two-electron dynamics are well described by the truncated basis. The
calculated photoionization spectrum is compared to the experimental one in
Fig.~\ref{fig:N12-13_spectra}(b). The resonance marked by the filled green bar is
indeed well reproduced by the theoretical calculation and the underlying
resonance wavefunctions [Fig.~\ref{fig:N12-13_spectra}(a)], similar to those
of Fig.~\ref{fig:wavefunctions}, confirm the attribution to the series
described in the main text.

\section{Adiabatic model potential}
The adiabatic potential~\eqref{eq:stark_potential} is obtained from the two-electron Hamiltonian~\eqref{eq:two-electron_hamiltonian}, neglecting dielectronic core polarization, in two steps. First, we assume that the motion of the two electrons can be adiabatically separated,
\begin{equation}
	\Psi(\bm{r}_1, \bm{r}_2) = \psi(\bm{r}_1 ; r_2)\phi(\bm{r}_2) ,
\end{equation}
with the wavefunction $\psi(\bm{r}_1 ;r_2)$ describing the motion of the faster inner electron depending parametrically on the radial position $r_2$ of the slower one --  much as in the Born-Oppenheimer approximation where the electronic wavefunction of a diatomic molecule parametrically depends on the internuclear distance. The function $\psi$ statisfies the equation
\begin{equation}
	\left(-\frac12\bm{\nabla}_1^2 + V_{l_1}(r_1) + \frac{1}{r_{12}}\right) \psi = \epsilon(r_2) \psi .
\end{equation}
The multipole expansion of the electronic repulsion reads
\begin{equation}
	\frac{1}{r_{12}} = \sum_q \frac{r_<^q}{r_>^{q+1}}P_q(\cos\theta_{12}) ,
\end{equation}
with $r_<$ ($r_>$) being the smallest (largest) of the two radial distances $r_1$ and $r_2$, respectively, and $P_q$ is the Legendre polynomial of order $q$. Keeping only the monopole and dipole terms ($q=0$ and $1$) and assuming that $r_2 > r_1$, a property easily verified by inspecting the probability densities in Fig.~\ref{fig:wavefunctions}, we obtain
\begin{equation}
	\left(-\frac12\bm{\nabla}_1^2 + V_{l_1}(r_1) + \frac{1}{r_2} + \frac{z_1}{r_2^2}\right) \psi = \epsilon(r_2) \psi , \label{eq:coree}
\end{equation}
with $z_1 = r_1\cos \theta_{12}$ the projection of $\bm{r}_1$ onto the adiabatic axis $\bm{r}_2$. Eq.~\eqref{eq:coree} describes the valence electron of the Sr$^+$ ion in the presence of an electric field of strength $1/r_2^2$ applied along the $z$ direction, and an additional constant energy offset $1/r_2$. Its eigenvalues are therefore
\begin{equation}
	\epsilon(r_2) = E_1^{\alpha,|\Lambda|}(r_2) + \frac{1}{r_2} ,
\end{equation}
with $E_1^{\alpha,|\Lambda|}(r_2)$ the energies of the Stark states of the Sr$^+$ ion (Fig.~\ref{fig:adiabatic_stark_potential}). Finally, we assume that, because the angular momentum of the outer electron is large, it does not penetrate much in the Sr$^{2+}$ core region and $V_{l_2}(r_2)$ is well approximated by the pure Coulomb potential $-2/r_2$. This approximation is justified by the fact that the quantum defects of the high-$l$ states of Sr$^+$ are very small~\cite{lange91}. The radial, adiabatic, one-electron Schrödinger equation for the \emph{outer} electron can now be written as
\begin{equation}
	\left(-\frac12\frac{\mathrm{d}^2}{\mathrm{d}r_2^2} - \frac{1}{r_2} + \frac{l_2(l_2+1)}{2r_2^2} + E_1^{|\Lambda|}(r_2)\right) \phi(r_2) = E \psi(r_2) ,
\end{equation}
which justifies the potential for the outer electron given in Eq.~\eqref{eq:stark_potential}. 

More quantitative adiabatic and nonadiabtic potentials for the outer electron
have been developped~\cite{braun90,richter92,grozdanov20a}. The present one is
simple and qualitative -- we do not expect the adiabatic approximation to hold
as well as it does in molecules -- but provides a remarkably clear physical
insight into the origin of the PPS and the electronic dynamics at play.

\end{document}